# Lateral ferromagnetic domain control in Cr$_2$O$_3$/Pt/Co positive exchange bias system

T. Nozaki[1,a], M. Al-Mahdawi[1], S. P. Pati[1], S. Ye[1], and M. Sahashi[1,2]

[1]*Department of Electronic Engineering, Tohoku University, Sendai 980-8579, Japan*

[2]*ImPACT Program, Japan Science and Technology Agency, Tokyo 102-0076, Japan*

We investigated the perpendicular exchange bias (PEB) switching from negative- to positive-exchange bias state for Cr$_2$O$_3$/Pt/Co exchange coupling thin film system exhibiting positive exchange bias phenomena. By changing Pt spacer layer thickness or measurements temperature, we demonstrated the control of two kind of intermediate state of the switching; the double hysteresis loop indicating local, non-averaged PEB, and single hysteresis loop indicating averaged PEB. We proposed the way to control the lateral ferromagnetic domain though the control of PEB magnitude.

---

[a] Author to whom correspondence should be addressed. –Electronic mail: nozaki@ecei.tohoku.ac.jp

Cr$_2$O$_3$/Co exchange coupling thin film system is an interesting system where a very large perpendicular exchange bias (PEB), $J_K$ > 0.3 erg/cm$^2$, is easily obtained.[1-3] Its magnitude can be controlled by inserting metal spacer between Cr$_2$O$_3$ and Co.[1-6] In additions, this system has received considerable attentions for the application of voltage controlled magnetic storage and memory devices, since an electric switching of the PEB by utilizing magnetoelectric (ME) effect of Cr$_2$O$_3$ by magnetoelectric field cooling (MEFC) process [3,7-10] and isothermal process[3,5,11-12] was demonstrated. Such a PEB switching was also achieved by only magnetic field cool (MFC) process with varied cooling magnetic field ($H_{fc}$) for Cr$_2$O$_3$/Pt/Co system,[2] which is called positive exchange bias phenomena.[13-16] In both PEB switching cases, multi-domain state of Cr$_2$O$_3$ antiferromagnetic (AFM) domain is observed after MFC/MEFC process in a certain cooling magnetic and electric field. During MEFC (MFC) process, competition between interface exchange coupling energy and magnetoelectric energy (Zeeman energy of Cr$_2$O$_3$ uncompensated moment) will occurs for MEFC case (positive exchange bias case). When the small magnetoelectric (magnetic) field was applied, magnetically aligned ferromagnetic (FM) domain state is imprinted to AFM domain via the exchange coupling, then one AFM domain state is stabilized. When the large field is applied, the magnetoelectric energy (the Zeeman energy) affect directly to AFM domain, then the different AFM domain state is stabilized. When the intermediate field is applied, AFM multi-domain state develop in consequence of these competitions. In such an intermediate field conditions, usually averaged PEB over the size of measurement region is observed; PEB magnitude decrease and become zero when the probability of two AFM domain of Cr$_2$O$_3$ (F$^+$: ↑ ↓ ↑ ↓ and F$^-$: ↓ ↑ ↓ ↑) become 50:50. However, sometimes double hysteresis loop with non-averaged (local) PEB is also observed. Such a double loop has been observed for positive exchange bias systems[15-16] and for "general" AFM/FM exchange bias systems with demagnetized state.[17-21] The double loop is explained in terms of the AFM and FM domain size. Only when the FM domain size is smaller than or comparable to the AFM domain size, such a local PEB is observed.[15] In case of $Cr_2O_3$/Co systems, a usual averaged PEB was mainly observed for Pt spacer systems,[2,8] while a local PEB was observed partly for Cr spacer system.[5] In additions, domain by domain coupling of Co and $Cr_2O_3$ surface spin was visually observed for no-spacer system by x-ray magnetic circular dichroism (XMCD) measurements,[3,22] which is example of a local PEB. Designing these two intermediate state is required with respect to each applications. In fact, the magnetic domain of $Cr_2O_3$ received increasing attentions toward device applications; Experimental demonstration of a local writing of exchange biased domain of $Cr_2O_3$/FM heterostructure,[23] theoretical study on domain wall dynamics,[24] and domain wall width[25] were reported recently. However, thus far, there's no attempt to control these two state actively. In this study, we investigated the intermediate state of the Pt-spacer samples exhibiting perpendicular exchange bias from its Pt spacer thickness dependence and temperature dependence, and discussed the role of exchange coupling for controlling these two state.

Samples were fabricated using a radiofrequency magnetron sputtering (RFMS) system. The film design was $Al_2O_3$ (0001) substrate/Pt (25 nm)/$Cr_2O_3$ (1000 nm)/Pt ($t_{Pt}$ nm)/Co (1 nm)/Pt (5 nm). The Pt spacer layer thickness $t_{Pt}$ was varied from 1.2 to 1.5 nm. The $Cr_2O_3$ film was grown in an Ar and $O_2$ mixture gas at 773 K. The Pt spacer layer thickness represented in this paper was determined from its deposition time where the sputtering rate is adjusted to be 0.048 nm/sec. Magnetization curves were measured by using a superconducting quantum interference device (SQUID) magnetometer or anomalous Hall effect (AHE), after a MFC process with varied $H_{fc}$ from 330 K. Slight difference in magnetic properties obtained by SQUID measurements and AHE measurements are observed, possibly due to the effect of the Hall bar microfabrication and small difference of temperature for the AHE measurements. The description of samples preparation and measurements set up were reported elsewhere.[5,8]

Fig. 1 (a) shows temperature dependence of PEB of samples with different $t_{Pt}$ measured by SQUID. $H_{ex}$ decrease with increasing Pt spacer thickness. $H_{ex}$ at 50 K change from 1,200 Oe to 200 Oe by increasing $t_{Pt}$ from 1.2 to 1.5 nm, while no spacer sample exhibit $H_{ex}$ ~ 3,500 Oe.[2,5] At around $t_{Pt}$ ~ 1.2 nm, $H_{ex}$ was very sensitive to the $t_{Pt}$; ~ 600 Oe change by only 0.1 nm change of $t_{Pt}$ was observed. Such a drastic change was also observed for our previous work,[2] while the absolute value of $t_{Pt}$ is slightly different possibly due to the different deposition conditions of Pt layer. The change of $H_{ex}$ possibly reflect the weakened Cr-Co coupling by increasing space between $Cr_2O_3$ and Co, which may be affected by the growth mode of Pt spacer layer. Fig. 1 (b) plots $H_{ex}$ value at 250 K after MFC with small $H_{fc}$ (+0.2 ~ +1 kOe) and large $H_{fc}$ (+5 ~ +7 kOe) measured by AHE. For all samples, we observed PEB switching from negative $H_{ex}$ (small $H_{fc}$) to positive $H_{ex}$ (large $H_{fc}$). By using a relative thick $Cr_2O_3$ thickness (1000 nm), we observed positive-$H_{ex}$ state for large variety of samples with relative small $H_{fc}$ < 7 kOe. Then we investigated intermediate state of PEB switching for all samples. Fig. 2 shows magnetization curves of characteristic samples; (a) Pt 1.3 nm sample and (b) Pt 1.2 nm sample measured by AHE at 250 K, after MFC with small, intermediate, and large $H_{fc}$. Samples with small $H_{ex}$ (Pt 1.3 ~ 1.5 nm) exhibit single hysteresis loop and averaged PEB, same as previous reports.[2,8] With increasing $H_{fc}$, $H_{ex}$ magnitude change from negative to positive though zero. On the other hand, in the case of Pt 1.2 nm sample, a two-step hysteresis loop were observed at intermediate $H_{fc}$. Since the hysteresis loop compose of both negative and positive $H_{ex}$ components, we regard it as a double hysteresis loop, while the two components are not well separate due to the small $H_{ex}$. Previous reports indicate the intermediated state can be changed by varying sample structure (spacer layer material). However in this report, we show it can be changed by controlling only PEB magnitude with keeping sample structure.

Next we checked the temperature dependence of the intermediate state for the Pt 1.2 nm sample. Interestingly, while a double hysteresis loop was observed at low temperature, the loop shape change with increase temperature, and near $T_N$ of $Cr_2O_3$, the loop become single. Fig. 3 compares $H_{fc}$ dependence of magnetization curves of Pt 1.2 nm sample measured by SQUID at (a) 280 K and (b) 50 K. The $H_{ex}$ and $<M_r>$ (average of upper y–intercept $M_r^+$ and lower y-intercept $M_r^-$ of the hysteresis loop) values, which is indicated by vertical and horizontal bars with arrows, are displayed in Fig. 3 (a) and (b). It is noted that we measured magnetization curve at 50 K and 280 K sequentially with keeping $T < T_N$, after a MFC with each $H_{fc}$. After MFC with $H_{fc}$ = 3.3 kOe, a double hysteresis loop with negative and positive $H_{ex}$ domain ratio almost 50:50 state was obtained at 50 K. On the other hand, at 280 K, single loop with $H_{ex} \sim 0$ was obtained. Smaller (larger) $H_{fc}$ make negative (positive) $H_{ex}$ ratio at 50 K increase, and absolute value of $H_{ex}$ at 280 K increase. Here we estimate the lateral AFM domain change against $H_{fc}$ at 280 K and 50 K. We define the volume fraction ratio $<F>$ as $<F> = (v^+ - v^-)/(v^+ + v^-)$, where $v^+$ and $v^-$ are volume ration of $F^+$ and $F^-$ domain. If we consider antiparallel coupling between Co and Cr surface spin,[26] negative (positive) $H_{ex}$ state correspond to $F^+$ ($F^-$) domain state. For a single loop case, $H_{ex}$ is represented by $J<S_{FM}><S_{AF}>/M_s t_{FM}$,[27-28] where J, $<S_{FM}>$, $M_s$, $t_{FM}$ are the exchange coupling constant, the averaged FM interface spin ($<S_{FM}> \sim -1$, since we assume antiparallel coupling between Co and Cr surface spin), the saturation magnetization of FM, and the thickness of FM. $<S_{AF}>$ is the averaged interface AFM spin and is same meaning as $<F>$. Thus $H_{ex}$ is proportional to $-<F>$. For an ideal double loop case such as Fig. 4 of Ref. 5, the negative-/positive-$H_{ex}$ domain ratio, that is <F>, can be represented by y-axis intercept of the hysteresis loop. In this study, because the hysteresis loops in Fig 3 (a) are not ideally separated, we regards <$M_r$> as the alternative. Figure 4 shows $H_{fc}$ dependence of (a) $H_{ex}$ at 280 K and (b) <$M_r$> at 50 K. The estimated <F> values are shown in the left axis of the figures. Error bars originate from measurement steps are shown in Fig. 4 (a). The $H_{fc}$ dependence of <F> in positive exchange bias system are well represented by a simplified equation taking into account the Zeeman energy of the surface spin of antiferromagnet as well as an antiferromagnetic exchange coupling at the AFM/FM.[2, 14]

$$\langle F \rangle \propto -\tanh\left(\frac{J_K}{k_B T_N} + \frac{g\mu_B}{k_B T_N} H_{fc}\right) = \tanh\left(\frac{g\mu_B}{k_B T_N} * \left(H_{fc} + \frac{J_K}{g\mu_B}\right)\right) \quad (1)$$

Fig. 4 (a) and (b) are fitted by $-A*\tanh(B*(H_{fc}+C))$ to compare the $H_{fc}$ dependence of <F>, assuming <F> proportional to $-H_{ex}$ and <$M_r$>. Fitting curves and obtained fitting parameters are also shown in Fig. 4. Fig 4 (a) and (b) are well fitted within its experimental error bars. It is difficult to estimate entirely correct <F> from Fig. 3 (a) and (b), because these are not ideal; In Fig. 3 (a), slight magnetization curve shape change was also observed in additions to the $H_{ex}$ magnitude change. In Fig. 3 (b), positive and negative $H_{ex}$ component are not fully separated. Considering these circumstances, the fitting parameters of (a) and (b) are agree very well. This agreement indicates the lateral AFM domain is stable during the temperature change from 50 K to 280 K, in spite of the drastic change of the lateral FM domain.

Here we consider the conditions to obtain the double hysteresis loop state in terms of the energy of the system. Figure 5 draw the schematic image of $Cr_2O_3$/Co interface when $Cr_2O_3$ is multi-domain state. For simplification, Pt spacer layer was not displayed in the figures. Primarily, Co-Co exchange interaction inside Co seek to align Co moments, while Co-Cr exchange coupling seek to imprint $Cr_2O_3$ domain to Co. Due to these competitive interactions, both state has higher energy and unstable compared with free standing Co. The single loop case (averaged PEB case), as indicated by red diagonal lines in Fig. 5 (a), there's parallel Co-Cr region, which align against exchange coupling. The energy of the system will increase by $J_K S$ due to the unfavorable Co-Cr connection, where S is the area of the unfavorable connection regions. On the other hand, the double loop case (local PEB case), as indicated by red diagonal line in Fig. 5 (b), domain wall develop in the Co layer, against the Co-Co exchange interaction. The energy of the system will increase by $\gamma l t_{Co}$, where $\gamma$ ($\sim \pi\sqrt{AK}$) is energy of domain wall formation, l is length of domain wall, $t_{Co}$ is thickness of Co (1 nm). From these considerations, double hysteresis loop will stabilized when the relations $J_K S > \gamma l t_{Co}$ is satisfied. Here, S and l are governed by $Cr_2O_3$ domain size, and the $Cr_2O_3$ domain size is less sensitive to the exchange coupling. In fact, as shown in Fig. 4, AFM domain is stable against temperature, in spite of large change of $J_K$ magnitude. Since $\gamma$ will not change largely at temperature much lower than its Curie temperature $T_C$, the increasing $J_K$ magnitude by using thinner spacer layer (Fig 2) or decreasing temperature (Fig 3) should responsible for the appearance of the double hysteresis loop. These results shows that we can achieve local, non-averaged PEB by increasing $H_{ex}$. In our experiments, we observed double hysteresis loop only when the $H_{ex}$ magnitude exceed the coercivity $H_c$. We believe the condition $H_{ex} > H_c$ become an indicator to satisfy the relations $J_K S > \gamma l t_{Co}$ and obtain local PEB, since previous reports[3,5,15-22] roughly satisfy this relation.

In conclusion, we investigated the intermediate state of PEB switching from negative- to positive-$H_{ex}$ state using $Cr_2O_3$/Pt/Co exchange coupling thin film system exhibiting positive exchange bias phenomena. Both the double loop (local PEB) and single loop (averaged PEB) in intermediate state can be realized by changing Pt spacer layer thickness or measurement temperature, even if its sample structure keep unchanged. Our results indicate the two intermediate state (local/ averaged PEB) can be controlled by $H_{ex}$ magnitude. Such a local/averaged PEB control would be essential for the design of each device applications.

This work was partly funded by JSPS KAKENHI Grant Number 16H05975 and ImPACT Program of Council for Science, Technology and Innovation (Cabinet Office, Japan Government).

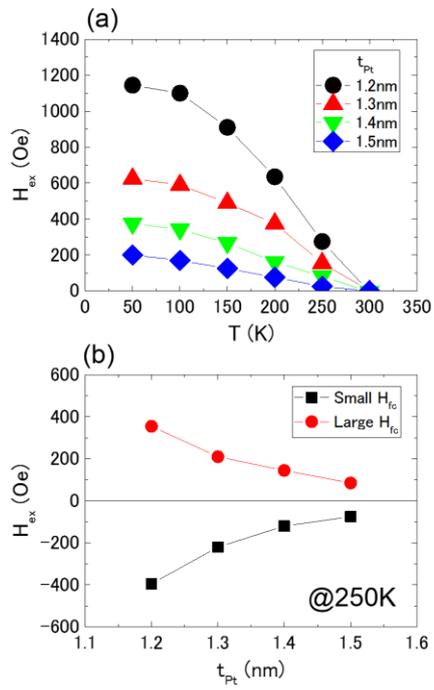

Fig. 1 (a) Temperature dependence of perpendicular $H_{ex}$ of samples with different Pt spacer layer thickness measured by SQUID. (b) Pt spacer layer thickness $t_{Pt}$ dependence of perpendicular $H_{ex}$ at 250 K measured by AHE after MFC with small and large $H_{fc}$.

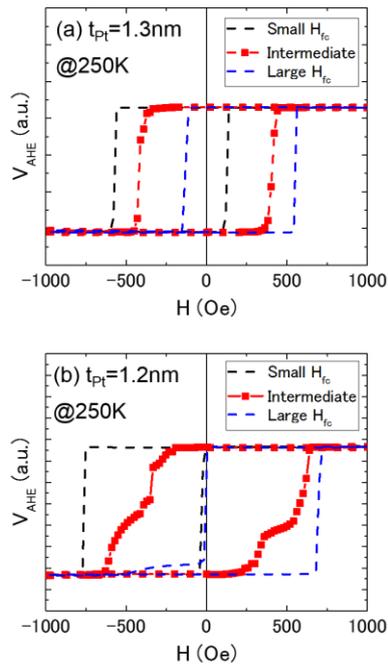

Fig. 2 Magnetization curves of samples with (a) $t_{Pt}$ = 1.3 nm (averaged PEB) and (b) $t_{Pt}$ = 1.2 nm (local PEB) measured by AHE after MFC with small, intermediate, and large $H_{fc}$.

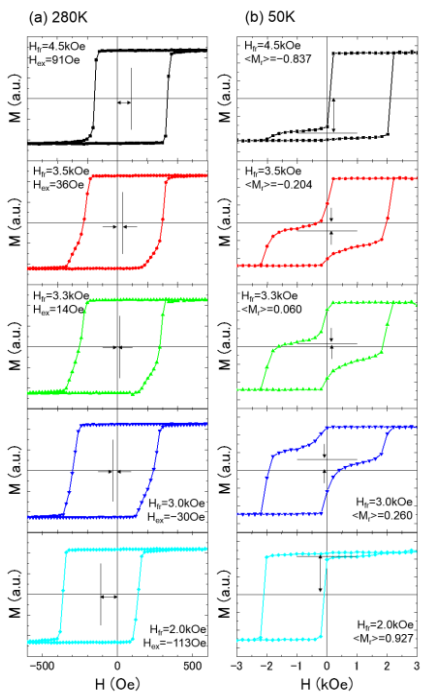

Fig. 3 Magnetization curves of $t_{Pt}$ = 1.2 nm sample at (a) 280 K and (b) 50 K measured by SQUID after MFC with different $H_{fc}$. The $H_{ex}$ value for (a) and average of upper y–intercept $M_r^+$ and lower y-intercept $M_r^-$ of the hysteresis loop, $<M_r>$ value for (b) are shown and indicated by vertical and horizontal lines in the figures.

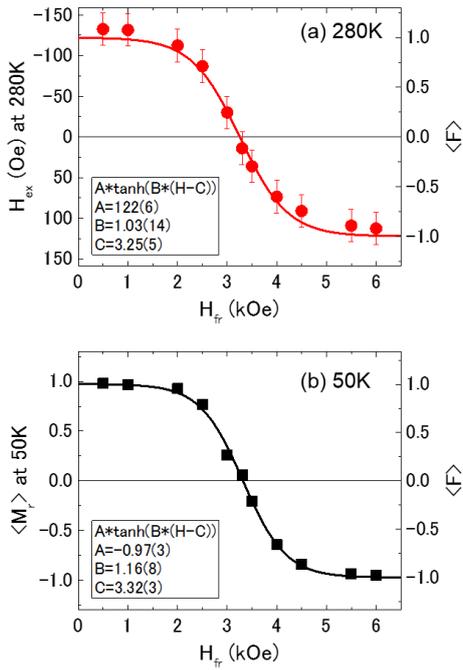

Fig. 4 $H_{fc}$ dependence of (a) the $H_{ex}$ value at 280 K and (b) the $<M_r>$ value at 50 K of $t_{Pt}$ = 1.2 nm sample obtained from Fig. 3. The left axis represent estimated $<F>$ values. Fitting curves and its fitting parameters are also shown.

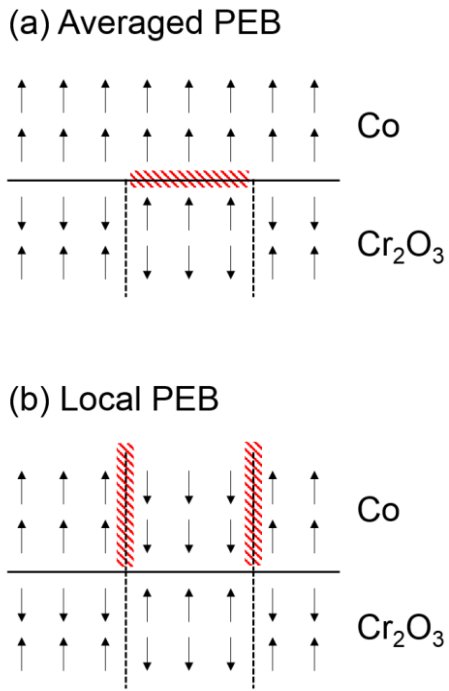

Fig. 5 Schematics of $Cr_2O_3$/Co interface spin of intermediate state of (a) averaged PEB case (correspond to Fig. 2 (a) and Fig. 3 (a)) and (b) local PEB case (correspond to Fig. 2 (b) and Fig. 3 (b)). Energetically unstable regions are indicated by red diagonal lines.